\documentclass[showpacs,aps,superscriptaddress,prb,twocolumn,amsmath]
{revtex4}
\usepackage{amssymb}
\usepackage{graphicx}
\usepackage{dcolumn}
\usepackage{bm}
\usepackage{hyperref}

\begin{document}

\preprint{APS/123-QED}

\title{Magnetic Properties of Single Transition-Metal Atom Absorbed Graphdiyne and Graphyne Sheet}
\author{Junjie He}
\affiliation{Department of Physics, Xiangtan University, Xiangtan
411105, China}
\author{ShuangYing Ma}
\affiliation{Department of Physics, Xiangtan University, Xiangtan
411105, China}
\author{Pan Zhou}
\affiliation{Department of Physics, Xiangtan University, Xiangtan
411105, China}
\author{C. X. Zhang}
\email{zhangchunxiao@xtu.edu.cn} \affiliation{Laboratory for Quantum
Engineering and Micro-Nano Energy Technology, Xiangtan University,
Xiangtan 411105, China}
\author{Chaoyu He}
\affiliation{Laboratory for Quantum Engineering and Micro-Nano
Energy Technology, Xiangtan University, Xiangtan 411105, China}
\author{L. Z. Sun}
\email{lzsun@xtu.edu.cn} \affiliation{Department of Physics,
Xiangtan University, Xiangtan 411105, China}

\date{\today}

\begin{abstract}
\indent The electronic and magnetic properties of single 3\emph{d} transition-metal(TM) atom (V, Cr, Mn, Fe, Co, and Ni) adsorbed graphdiyne (GDY) and graphyne (GY) are systematically studied using first-principles calculations within the density functional framework. We find that the adsorption of TM atom not only efficiently modulates the electronic structures of GDY/GY system, but also introduces excellent magnetic properties, such as half-metal and spin-select half-semiconductor. Such modulation originates from the charge transfer between TM adatom and the GDY/GY sheet as well as the electron redistribution of the TM intra-atomic \emph{s, p,} and \emph{d} orbitals. Our results indicate that the TM adsorbed GDY/GY are excellent candidates for spintronics.\\
\end{abstract}
 \maketitle
\section{Introduction}
\indent Graphene,\cite{1} a single atomic layer of carbon sheet, has attracted tremendous attention because of its unique properties such as high room temperature carrier mobility\cite{2,3} and quantum Hall effect.\cite{4,5} Graphyne (GY), a two dimensional carbon allotrope with the same symmetry as graphene predicted to have a high possibility of synthesis by Baughman \emph{et al.} in 1987\cite{6}, is made up of hexagonal carbon rings and acetylene linkages.\cite{7} So far, considerable experimental efforts have devoted to synthesize such unusual materials and numerous of related polymer structures have been prepared successfully.\cite{8,9,10,11} Graphdiyne (GDY), belonged to the family of GY, is also mesh structure consisting of carbon hexagonal rings and butadiyne linkages.\cite{12,13,14} Most intriguingly, (GDY) has been successfully grown on the surface of copper via a cross-coupling reaction using hexaethynylbenzene.\cite{15,16} Consequently, the preparation of GY can be expected because theoretical work has shown that it is energetically more stable than GDY. In comparison with graphene, GY and GDY possess many interesting properties because they posses both $sp$-hybridized and $sp^{2}$-hybridized carbon atoms and the natural 'holes'. For example, GY is a promising candidate for the anode material in lithium battery applications because of its higher lithium mobility and higher lithium storage capacity than graphite.\cite{17} On the other hand, the Ca decorated GY is predicted an excellent candidate for hydrogen storage because of its high capacity hydrogen storage and effective prevention for the formation of Ca cluster.\cite{18}\\
\indent Analogous to the graphene, GDY/GY possess small carrier effective masses,\cite{12,19} high carrier mobilities.\cite{13} In particular, based on density functional theory (DFT) calculations the GY and GDY are found to be semiconductors with direct band gap of 0.52eV and 0.53eV, respectively,\cite{12} which makes them advanced in preparation of effective room temperature field-effect transistors(FET). To expand the application of GDY/GY, modulation of their electronic properties is essential. Recent studies have indicated that similar to that of graphene,  GDY/GY nanoribbons can effectively tailor the electronic properties of GDY/GY.\cite{13,20} Interestingly, new families of carbon nanotubes based on graphyne motifs with even richer variation in electronic properties than do ordinary single wall nanotubes were predicate by Coluci et al\cite{tube1, tube2}. Doping is another effective approach to modulate the electronic properties of GDY/GY. For example, GY with adsorption of Ca represents spin-polarized electronic property with a small localized magnetic moment($\sim$0.25$\mu_B$).\cite{18} It has been proved that the surface adsorption of TM atom profoundly influences the electronic properties of graphene and induces spin polarization\cite{21,22,23,24,25,26} in the system, which makes is a vital candidate materials for spintronics.\cite{27,28} However, to our knowledge, the effects of adsorption of TM atom on the surface of  GY/GDY has not been reported as yet. It is significant to study such effect for their promising applications. To this end, we systematically investigate the electronic and magnetic properties of GDY and GY with single TM (V, Cr, Mn, Fe, Co and Ni) atom adsorption (TM-GDT/GY) using the first-principles calculations within the framework of the density functional theory (DFT). We find that the adsorption of TM atom not only efficiently modulates the electronic structures of GDY/GY system, but also introduces excellent magnetic properties, such as half-metal and spin-select half-semiconductor.\\
\section{METHOD AND CALCULATION DETAILS}
\begin{figure}
 \includegraphics[width=3.3in]{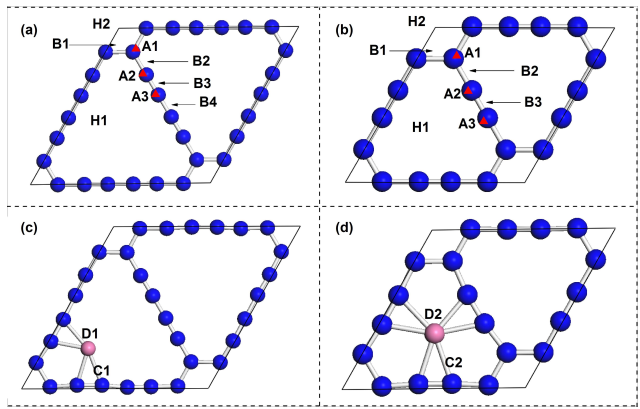}
 \caption{Diagrams of TM-GDY/GY. (a) and (b) are the structures of pure GDY and GY, respectively. (c) and (d) illustrate the most stable structures of TM-GDY and TM-GY, respectively. Possible adsorption sites are marked in (a) and (b), where $A_{i}$ ,$B_{i}$ and $H_{i}$ denote top, bridge and hollow sites,
 respectively. Blue and pink balls represent the carbon and TM atoms, respectively.}
 \label{fig1}
\end{figure}

\indent A super-cell of GY/GDY with TM adatom is adopted to be the calculated model as shown in Fig.\ref{fig1}, where a vacuum space of 13{\AA} perpendiculars to the GY/GDY plane is chosen to avoid the interactions between the neighboring images. The first-principles calculations are performed using the Vienna ab initio simulation package (VASP)\cite{29, 30} within spin-polarized DFT.\cite{31,32} As reported by Wehling et al.\cite{33} that the electronic configuration, the adsorption geometries, and the magnetic state of the 3$d$-TM adatom on graphene are very sensitive to the effects of local Coulomb interactions U in the TM $d$ orbitals. We employed a generxalized gradient approximation functional (GGA)\cite{34} as well as GGA + U to describe the exchange-correlation functional. We use U = 4 and J=0.9 eV to study the dependence of the electronic structure on U, which is a typical value for 3$d$ TM.\cite{35,36}  The projected augmented wave (PAW)\cite{37,38} potentials are adopted to describe the interaction between nucleus and valence electrons. A plane-wave basis set with the kinetic energy cutoff of 400 eV is employed. The Brillouin zone (BZ) is sampled using $5\times5\times1$ and $9\times9\times1$ Gamma-centered Monkhorst-Pack grids for the calculations of relaxation and electronic structures, respectively. All systems are fully relaxed without any symmetric constrains. The criteria of energy and atom force convergence are set to $10^{-5}$ eV/unitcell and 0.01 eV/{\AA}, respectively. The dipole correction\cite{39} is considered to deal with the impact of variety of potential distribution introduced by TM adsorption. All calculation parameters in present work are systematically optimized.\\
\indent To evaluate the most favorable adsorption configuration, adsorption energy E$_a$ is adopted which is defined as,
\begin{displaymath}
E_{a}=E_{TM+sheet}-E_{sheet}-E_{TM}
\end{displaymath}
where $E_{TM+sheet}$ denotes the spin-polarized total energy of TM adsorbed GY/GDY sheet, $E_{sheet}$ is the total energy of the isolated pure GY/GDY sheet, and $E_{TM}$ is the spin-polarized total energy of the corresponding free TM atom.\\
\indent To analyze spin polarization of the systems, we define the spin polarization $P(E_{(f/HOMO)})$ at the Fermi level or at the highest occupied molecular orbital (HOMO) as:
\begin{displaymath}
P(E_{(f/HOMO)})=\frac{D(E_{(f/HOMO)\uparrow})-D(E_{(f/HOMO)\downarrow})}{D(E_{(f/HOMO)\uparrow})+D(E_{(f/HOMO)\downarrow})}
\end{displaymath}
where $D(E_{(f/HOMO)\uparrow})$ and $D(E_{(f/HOMO)\downarrow})$ represent the DOS of the majority spin and minority spin at the Fermi level or the HOMO state, respectively.\\
\indent The charge transfer between the TM adatom and the GDY/GY sheet are evaluated using the atomic basin charge based on the atoms in molecular (AIM) theory\cite{40}. The AIM theory gives a unique definition of partitioning space into atomic basins by simply knowing the total charge density. The atomic basin is defined as a region of space traversed by trajectories of the electron density gradient terminating at a given nucleus (attractor) and enclosed inside a zero charge-density flux surface $\overrightarrow{\nabla}\rho(\vec r)\cdot\overrightarrow{n}(\vec r)=0$. Here $\overrightarrow{n}(\vec r)$ is the unit vector of the surface. Thus, space is divided into regions called atomic basin by the surfaces that run through the points where $\overrightarrow{\nabla}\rho(\vec r)=0$. By integrating the electronic density within the basin, the charge of each atom can be estimated.\\
\indent To have a better understanding of the magnetism of the TM-GDY/GY system, we use the map of spin-polarized charge density(SCD) for all investigated systems. The SCD are defined as:
\begin{displaymath}
\rho(\vec r)=\rho\uparrow(\vec r)-\rho\downarrow(\vec r)
\end{displaymath}
where $\rho\uparrow(\vec r)$ and $\rho\downarrow(\vec r)$ are the spin up and the spin down charge density of TM-GDY/GY system, respectively.\\
\begin{figure}
 \includegraphics[width=3.5in]{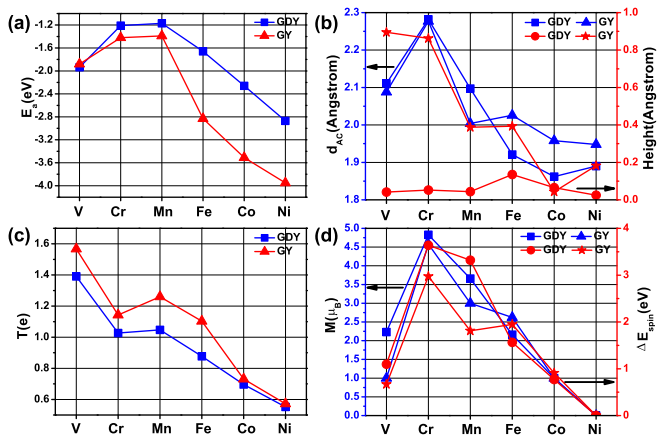}
 \caption{ (a) is the adsorption energy of TM-GDY/GY systems, (b) is the distance between TM adatom and its nearest carbon atoms on the GDY/GY
 sheet and height between the adatom and GDY/GY sheet. Charge transfer (T(e)) from the TM atom to GDY/GY sheet based on AIM theory is shown in (c), and (d) is the magnetic moment (M) and spin-polarized energy $\Delta E_{spin}$.}\label{fig2}
\end{figure}
\section{Results and discussions}
\subsection{Adsorption geometry and energy}
\indent Before investigating the TM adsorbed systems, we firstly optimize the pristine GDY and GY. The optimized lattice constants of GDY and GY are 9.50{\AA} and 6.99{\AA}, respectively, which is in good agreement with previous reports\cite {12,13,14}. In present work, six 3d TMs (V, Cr, Mn, Fe, Co and Ni) are selected to adsorb on the GDY and GY sheet. Considering the symmetry of GDY and GY, all possible initial adsorption sites are considered for each types of TM adatom, as shown in Fig.\ref{fig1} (a) and (b). We find that the most favorite adsorption site is $D_1$ for GDY and $D_2$ for GY for all the six 3d TMs, as shown in Fig.\ref{fig1} (c) and (d). All the TM adatoms disturb the lattice configuration of GDY/GY to some extent. Figure \ref{fig2} (a) and (b) give the TM adsorption energies and structure parameters for TM-GDY/GY, respectively. From Fig.\ref{fig2} (a), one can see that the considerable negative adsorption energies for all the 3d TM adatoms on  GDY/GY indicating their chemisorption characteristics. The largest adsorption energy is -1.17eV (-1.39eV) when the Mn adsorbs on the GDY(GY) sheet; the minimum one is -2.87eV (-3.95eV) when the Ni adsorbs on the GDY (GY) sheet. From Fig.\ref{fig2} (b), we find that as for TM-GDY the maximum (minimum) distance between the adatom and the nearest carbon atom ($d_{AC}$) is 2.282{\AA} (1.862{\AA}) for Cr (Co); and the maximum (minimum) height between adatom and GDY sheet is 0.135{\AA} (0.025{\AA}) for Fe (Ni). As for GY, the maximum (minimum)  $d_{AC}$ is 2.277{\AA} (1.948{\AA}) for Cr (Ni), which is similar to the case of GDY; whereas the maximum (minimum) height is 0.895{\AA} (0.043{\AA}) for V (Co). According to previous reports\cite{21,22,23,24,25,26}, the most favorable adsorption site for TM on graphene can be top, bridge and hollow site depending on the balance the contributions to chemical bonding from the s against those from the d orbitals of TMs\cite{33}, however, the most favorable adsorption site for TM on GDY/GY is all in the alkyne ring. Moreover, the adsorption energies for the TM-GDY/GY are much smaller than those of graphene, which indicates that the adsorption of TM atoms on the GDY/GY is more stable than the corresponding TM adsorption on graphene. Such phenomena originate from the different hybridization of carbon atoms between graphene and GDY/GY. There are only $sp^2$ carbon atoms in the graphene but $sp$ and $sp^2$ carbon atoms in the GDY/GY. The $sp$ carbon atoms play a key role in the absorption process of TM atoms on the GDY/GY. Besides the $sp^2$ hybridization as in the case of graphene, which only exists out-of-plane $p_z$ $\pi$/$\pi^{\ast}$ (here z is the direction perpendicular to the carbon plane), there are in-plane $p_x$-$p_y$ $\pi$/$\pi^{\ast}$ in GDY or GY for $sp$ hybridization. Such hybridization nature of GDY or GY enables the $\pi$/$\pi^{\ast}$ to rotate toward any direction perpendicular to the line of -C$\equiv$C-, thus making it possible for the $\pi$/$\pi^{\ast}$ states from the -C$\equiv$C- bonds at a given alkyne ring point toward the TM atom. As a result, the valance electrons of the TM atom couples with not only the $p_z$ but also the $p_x$ and $p_y$ of carbon atoms in the -C$\equiv$C-, which produces the TM atoms adsorb in the alkyne ring with more favorable energy. Moreover, the heights between TM adatom and the GDY/GY sheet are all smaller than that of graphene sheet, which originates from both the bigger size of acetylenic and butadiynic ring and the rotation of $\pi$/$\pi^{\ast}$ states of -C$\equiv$C-.\\
\begin{table*}
\caption{The calculated results of TM-GDY/GY. "M","HM", "HSM" and "SM" in the band structures (BS) column denote metal, half-metal, half-semiconductor and semiconductor, respectively. $T_e$ is the charge transfer from the TM to GDY and GY derived from AIM. "4$s^{\ast}$/3$d^{\ast}$/4$p^{\ast}$" denotes the the valence
electron configuration corresponding to the Bader charge analysis of the TM-GDY/GY. TM$\uparrow$/TM$\downarrow$ is the electron numbers of spin up and spin down for TM atomic basin. Magnetic moment of TM-GDY/GY system $\mu_T$($\mu_B$) and magnetic moment of TM atomic basin $\mu_a$($\mu_B$) in TM-GDY/GY system are also given. $\Delta E_{spin}$ is the energy difference defined as $\Delta E_{spin}$=$E_{NM}$-$E_M$, where $E_{NM}$ and $E_M$ are the total energies for non-magnetic and magnetic states in one unit cell, respectively.}\label{tab1}
\begin{ruledtabular}
\begin{tabular}{cccccccc}
\hline
 & BS &$T_e$  &$4s^*/3d^*/4p^*$&
 TM$\uparrow$/TM$\downarrow$  & $\mu_T$ & $\mu_a$ &
 $\Delta E_{spin}$(eV)\\
\hline
V-GDY &HM &1.391  &0.269/3.187/0.153  &3.219/0.390 &2.23 &2.83 &1.097\\
Cr-GDY&M  &1.027  &0.402/4.290/0.281  &4.388/0.586 &4.83 &3.80 &3.639\\
Mn-GDY&M  &1.047  &0.234/5.096/0.265  &4.483/1.113 &3.72 &3.37 &3.320\\
Fe-GDY&M  &0.877  &0.417/6.243/0.463  &5.005/2.119 &2.16 &2.89 &1.564\\
Co-GDY&M  &0.696  &0.494/7.324/0.517  &4.369/3.935 &1.00 &0.43 &0.771\\
Ni-GDY&SM &0.552  &0.388/8.677/0.383  &4.724/4.724  &0    &0    &0\\
\hline
V-GY  &HM &1.567  &0.348/2.558/0.527  &2.286/1.148  &1.00 &1.14 &0.670\\
Cr-GY &M  &1.143  &0.394/4.102/0.361  &4.427/0.429  &4.58 &4.00 &2.975\\
Mn-GY &HSM&1.260  &0.304/4.911/0.525  &4.203/1.537  &3.00 &3.05 &1.811\\
Fe-GY &M  &1.103  &0.451/5.819/0.626  &4.540/2.357  &2.62 &2.18 &1.949\\
Co-GY &HM &0.731  &0.607/6.871/0.819  &4.742/3.555  &1.00 &0.19 &0.910\\
Ni-GY &SM &0.572  &0.418/8.026/0.984  &4.714/4.714  &0    &0    &0\\
\hline
\end{tabular}
\end{ruledtabular}
\end{table*}
\begin{figure}
 \includegraphics[width=3.3in]{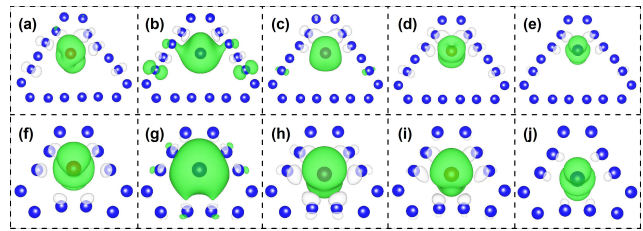}\\
 \caption{Spin-polarized charge density (SCD) distribution for V (a) , Cr (b), Mn (c), Fe (d) and Co (e) adsorbed GDY; V (f), Cr (g), Mn (h), Fe (i) and Co (j) adsorbed GY. The isosurface charge density is set to 0.004e/{\AA$^3$}.} \label{fig3}
\end{figure}
\subsection{Magnetic and Electronic structures of TM-GDY/GY}
\begin{figure}
 \includegraphics[width=3.5in]{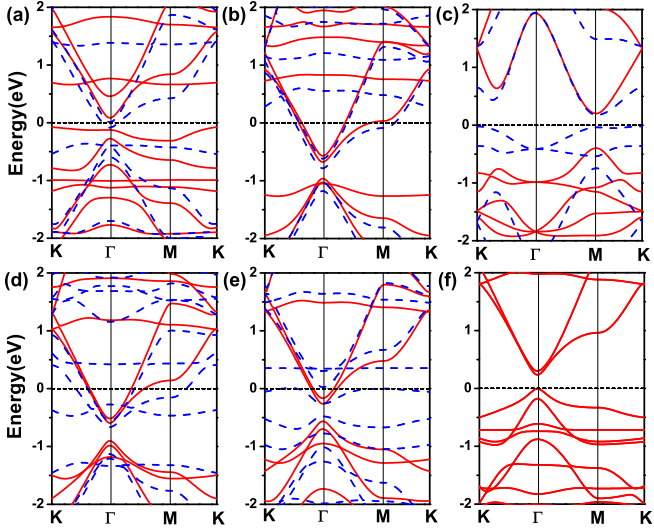}\\
 \caption{Energy band structures of V-GDY (a), Cr-GDY (b), Mn-GDY (c), Fe-GDY (d), Co-GDY (e), and Ni-GDY (e). The solid (red) and dash (blue) lines represent the majority and minority spin channels, respectively. The Fermi level is set to zero.}
 \label{fig4}
\end{figure}

\begin{figure}
 \includegraphics[width=3.5in]{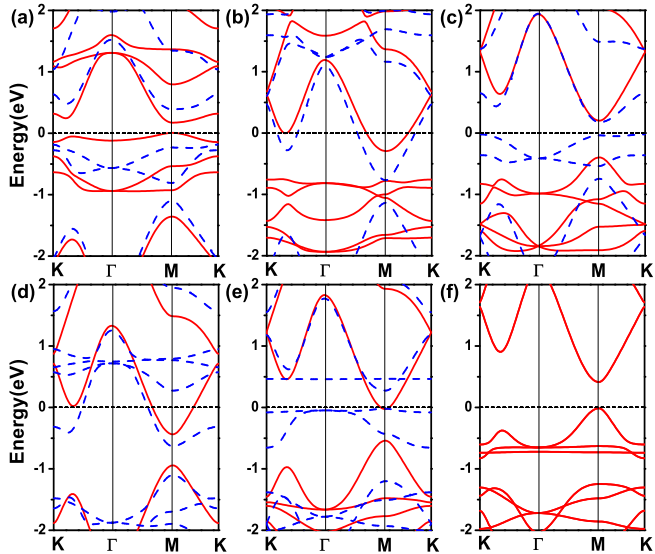}\\
 \caption{Energy band structures of  V-GY (a), Cr-GY (b), Mn-GY (c), Fe-GY (d), Co-GY (e), and Ni-GY (e). The solid (red) and dash (blue) lines represent the majority and minority spin channels, respectively. The Fermi level is set to zero.}
 \label{fig5}
\end{figure}

\begin{figure}
 \includegraphics[width=3.5in]{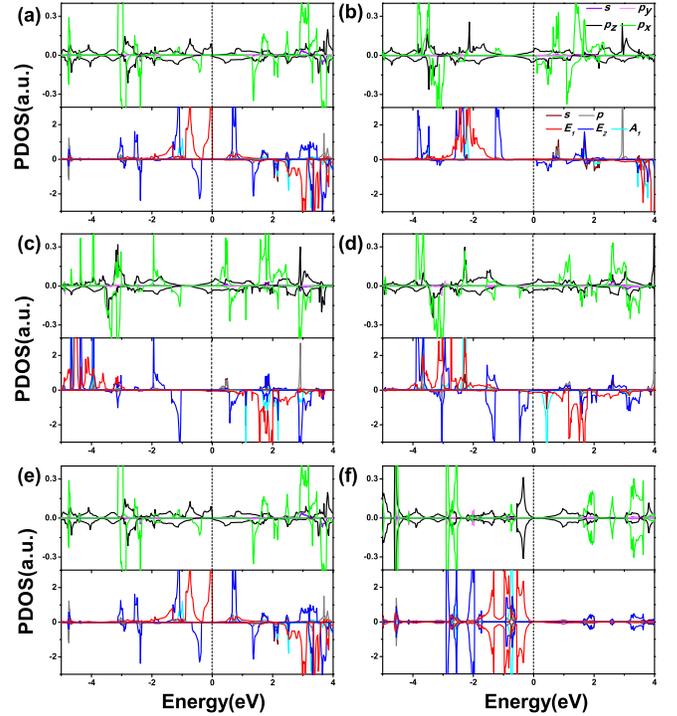}\\
 \caption{PDOS of V (a), Cr (b), Mn (c), Fe (d) , Co (e) and Ni (f) adsorbed GDY. The upper and lower panels are the PDOS of nearest neighbor carbon atom (C1) (as indicated in the \ref{fig1} (c)) of TM adatom and TM adatom, respectively. The Fermi level is set to zero.}
 \label{fig6}
\end{figure}

\begin{figure}
 \includegraphics[width=3.5in]{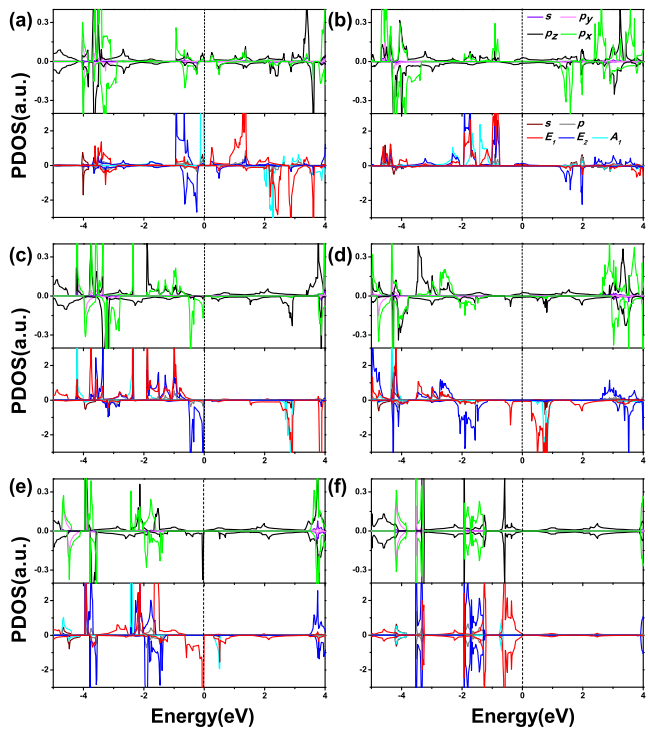}\\
 \caption{PDOS of V (a), Cr (b), Mn (c), Fe (d) , Co (e) and Ni (f) adsorbed GY. The upper and lower panels are the PDOS of nearest neighbor carbon atom (C2) (as indicated in the Fig.\ref{fig1} (c)) of TM adatom and TM adatom, respectively. The Fermi level is set to zero.}
 \label{fig7}
\end{figure}

\indent Based on the most stable structures of TM-GDY/GY, we systematically investigate their magnetic and electronic properties. Distinct spin polarization is found for the GDY/GY with adsorption of V, Cr, Mn, Fe and Co. The corresponding magnetic moment is given in Fig.\ref{fig2} (d) and Tab.\ref{tab1}. The magnetism of TM-GDY/GY is mainly contributed by the TM adatoms. The sequence of magnetic moment of TM-GDY/GY is Cr$>$Mn$>$V$>$Fe$>$Co for TM-GDY and Cr$>$Mn$>$Fe$>$V$=$Co for TM-GY, respectively. In comparison with the free-standing states of TM atoms, the magnetic moments of V, Cr, Mn, Fe and Co atom are all reduced when they adsorb on both GDY and GY. The reduction is derived from the electron injection from TM to GDY/GY and electron redistribution between the electron orbitals of TM induced by the strong coupling between TM and GDY/GY, which reduce the number of spin-polarized electrons of TM atom and produce the TM adatom in low-spin configuration, the details will be discussed below. In Fig. \ref{fig2} (d) and Tab.~\ref{tab1}, we also show the spin-polarized energy ($\Delta E_{spin}$) which is the energy difference between the non-magnetic state and the magnetic state. All the $\Delta E_{spin}$s are larger than 0.7 eV, indicating the stability of the spin-polarized states of GDY/GY with adsorption of this five TM atoms, and some of them are hoping to be high temperature FM systems. The sequence of $\Delta E_{spin}$ is similar to that of magnetic moment of the systems as a whole. For example, the Cr-GDY possesses the largest $\Delta E_{spin}$ corresponding to the largest magnetic moment. Spin-polarized charge density (SCD) as shown in Fig. \ref{fig3} represents that the coupling  between TM atom (expect Ni atom) and GDY or GY is antiferromagnetism and the magnetic moment of TM-GDY/GY is mainly contributed by the TM adatom. Moreover, one can see that much stronger SCD locates around the adatom for Cr and Mn adsorbed GDY (GY), consistent with their larger magnetic moment.\\
\indent The spin-resolved band structures of TM-GDY and TM-GY are shown in Fig. \ref{fig4} and Fig. \ref{fig5}, respectively. Our results indicate that the pristine GDY and GY are direct band gap semiconductors with band gap of  0.49eV located at the $\Gamma$ point and 0.48 eV located at the M point, respectively. The results are in good agreement with previous work\cite{12}. The adsorption of V, Cr, Mn, Fe and Co induces spin splitting of the band structures and produces the magnetism for TM-GDY/GY. However, Ni-GDY/GY shows no spin polarization and magnetism, which is consistent with the case of TM-adsorbed graphene\cite{21, 25}. Interestingly, versatile electronic characteristics are achieved in TM-GDY/GY. We summarize the magnetic and electronic properties of TM-GDY/GY as follows:\\
(i) Half-metal: V-GDY, V-GY and Co-GY behave as half-metal which can be used as spin filter. Their band structures are given in Fig. \ref{fig4} (a), Fig. \ref{fig5} (a) and Fig. \ref{fig5} (e), respectively. The V-GDY behaves as metal for majority spin channel and a semiconductor with an indirect band gap of 0.16eV for majority spin channel. The V-GY is metallic for majority spin channel and a semiconductor with an indirect band gap of 0.478 eV for minority spin channel. However, the Co-GY is a semiconductor with a direct band gap of 0.426 eV for majority spin channel and behaves as a metal for minority spin channel.\\
(ii) Spin-select half-semiconductor: The band structures of Mn-GY as shown in Fig. \ref{fig5} (c) indicate its spin-select half-semiconductor with 100$\%$ spin polarization at HOMO state, where both the maximum of the valence band (VBM) and minimum of the condition band(CBM) belong to the same spin channel. The majority spin channel of Mn-GY is direct band gap semiconductor with band gap of 0.602eV at the M point. However, the minority spin channel of Mn-GY is indirect band gap semiconductor with band gap of 0.20eV at the K point.\\x
(iii) Semiconductor-metal transition: The GDY and GY are spin-degenerated direct-gap semiconductors. However, when GDY is adsorbed by Cr, Mn, Fe and Co, as shown in Fig.\ref{fig4} (b), (c), (d) and (e), and GY is adsorbed by Cr and Fe, as shown in Fig.\ref{fig5} (b) and (c), the systems realize semiconductor-metal transition. All these systems are spin polarized metals. In particular, the spin polarization of Co-GY is up to 90$\%$.\\
(iv) Narrow-gap semiconductor: Although Ni does not induce spin polarization in GDY and GY, as shown in Fig.\ref{fig4} (f) and Fig.\ref{fig5} (f), it narrows the band gaps of GDY and GY to 0.24eV and 0.40eV, respectively. The k point in reciprocal space of VBM and the CBM is the same as the pristine GDY/GY sheet.\\
\indent All the above results indicate that the adsorption of TM on GDY/GY is an effective modulation approach on their magnetic and electronic properties. Such modulation mainly originates from the electron transfer between TM adatom and the GDY/GY, as well as between the TM intra-atomic orbitals induced by the strong coupling between TM and GDY/GY. To quantitatively investigate the modulation mechanism, we calculate the electron transfer for each adatom on the basis of Bader charge analysis.\cite{40} The electrons occupied 4\emph{s}, 3\emph{d}, and 4\emph{p} orbital of TM are obtained by separating total electrons in TM atomic basin based on the proportion of 4\emph{s}, 3\emph{d}, and 4\emph{p} partial density of states (PDOS) (as shown in Fig.\ref{fig6} and Fig.\ref{fig7}) to total density of states (TDOS). Obvious charge transfer takes place from the TM adatom to GDY/GY sheet as shown in Fig.\ref{fig2} (c) and Tab.\ref{tab1}. For both GDY and GY, the maximum and minimum electron transfer occurs when the systems are adsorbed by V and Ni, respectively. From Tab.\ref{tab1} we find that the electrons not only transfer from TM to GDY/GY but also redistribute between the \emph{4s, 3d} and \emph{4p} orbitals of TM adatom. The \emph{4s} orbital of TM adatoms always lose electrons and the \emph{4p} orbital always get electrons. As for GDY system, the 3\emph{d} orbital of V and Cr loses electrons, while the 3\emph{d} orbital of other TMs gains electrons. However, as for GY system,  only the 3\emph{d} orbital of Co and Ni trivially gain electrons. Such charge transfer will affect the pairing of electrons of TM and further influence the magnetic properties of TM-GDY and TM-GY. The results in Tab.\ref{tab1} show that the electrons occupied \emph{4s} orbital of TM adatom for all TM-GDY/GY systems are all less than 0.6 e. Such orbital occupancy will produce the TM adatom on GDY/GY in the low-spin configurations as like the case of TM adsorbed graphene reported by Wehling et al.\cite{33}.\\
\indent Such electron transfer is derived from the strong coupling between TM and GDY/GY which is determined by the bonding characteristics of GDY/GY. The $p_x$ and $p_y$ of \emph{sp} hybridized carbon atoms in the GDY and GY contribute to both $\sigma$-bond and $\pi$-bond, making it easy to couple with the TM adatom. The projected density of states (PDOS) of TM-GDY/GY systems is shown in Fig.\ref{fig6} and Fig.\ref{fig7} to illuminate the coupling between the TM and the GDY/GY sheets. We only show the PDOS for one of the NN \emph{sp} hybridized carbon atoms of the TM adatom in Fig.\ref{fig6} and Fig.\ref{fig7} because the TM adatoms only strongly bind to the nearest neighbor (NN) \emph{sp} hybridized carbon atoms and the NN \emph{sp} hybridized C atoms of the TM adatom in GDY or GY exhibit the similar results. Here we take Fe-GDY as example to elucidate the relationship between the coupling, electron transfer, and the magnetic properties of TM-GDY/GY. The PDOS of Fe adatom and C1 for Fe-GDY are shown in Fig.\ref{fig6} (d). For the majority spin, the Fermi level passes through the conduction band due to the electron transfer from TM 3\emph{d} to GDY \emph{p} and the states around the Fermi level are mainly derived from C1 $p_z$ states. Moreover, in the majority spin channel, the Fe $d$ orbital with E1 ($d_{yz}, d_{xz}$) and E2 ($d_{xy}, d_{x^2-y^2}$) symmetry strongly couples with the $p_x$ and $p_z$ of C1 in the energy ranges of -4.0$\sim$-3.4eV and -2.7$\sim$-1.0eV. For the minority spin, the states around the Fermi level are mainly derived form the coupling between Fe E1 and C1 $p_z$. One can see that there are considerable overlap for the Fe E2 and C1 $p_x$ and $p_z$ in -3.5$\sim$-2.9eV and -1.6$\sim$-1.2eV energy windows. However, from the electron statistics based on the Bader Charge analysis\cite{40} as listed in Tab.\ref{tab1}, we find that though the Fe atom injects 0.877e electrons into GDY, the total electrons of Fe 3\emph{d} is still increased by 0.243e, which is derived from the electron promotion from Fe 4\emph{s} to Fe 3\emph{d}. The intra-atomic electron transfer of Fe adatom is mainly determined by the coupling effect between the 4\emph{p} and 4\emph{s} of Fe and C1 $p_x$ and $p_z$ around -2.3eV, the coupling between the 4\emph{p}, E1, and  E2 of Fe and C1 $p_x$ and $p_z$  around -1.4eV in the majority spin channel, as well as the coupling between the 4\emph{p},  E1 and  E2 of Fe and $p_x$ and $p_z$ of C1 around -3.0eV in the minority spin channel. Along with the coupling between Fe and GDY, the empty Fe 4\emph{p} show 0.463e electron occupation. This is the result of electron exchange from GDY to the Fe adatom and the electron redistribution between intra-atomic orbitals from 4\emph{s} and 3\emph{d} to the 4\emph{p} for Fe adatom. It can be found that the states of Fe 4\emph{p} and C1 $p_x$ couple with each other around -3.7eV, -3.0eV and -1.4eV in both majority and minority spin channels as shown in Fig.\ref{fig6} (d), which corresponds to the electron exchange from the C1 $p_x$ to the Fe \emph{4p} states. Such electron exchange between Fe and GDY as well as the intra-atomic oribtal electron redistribution consequently affect the pairing of electrons of Fe and further determine the magnetic property of Fe-GDY. As a result, the Fe adatom behaves a low-spin configuration with 2.16$\mu_B$ magnetic moment in Fe-GDY system. The same physical mechanism results in the low-spin configurations of TMs in the TM-GDY/GY when GDY/GY is adsorbed by V, Cr, Mn, Fe, Co and Ni. \\
\indent Here, we summarize the electronic coupling characteristics near the Fermi level for TM-GDY/GY in details as follows:\\
(i) Half-metal: The PDOS of half-metals V-GDY, V-GY and Co-GY are shown in Fig.\ref{fig6} (a), Fig.\ref{fig7} (a) and Fig.\ref{fig7} (e), respectively. The V-GDY is metallic with the Fermi level passing through the conduction band determined by the states of $p_z$ of C1 in the minority spin channel. As for the majority spin channel, the VBM is derived from the coupling of V E1 and C1 $p_z$, as shown in Fig.\ref{fig6} (d). Fig.\ref{fig7} (a) indicates that the Fermi level of V-GY passes through the valence band in the majority spin channel where the VBM derive from the coupling between $p_x$ and $p_z$ of C2 and A1 ($d_{z^2}$) of V. However, in the minority spin, the VBM derives from the coupling between $p_x$ and $p_z$ of C2 and E2 of V and the CBM derives from the coupling between $p_x,p_z$ of C2 and E1 of V. As shown in Fig.\ref{fig7} (e), the Co-GY is a half-metal with the Fermi level passing through the valence band in minority spin channel which is mainly contributed by the coupling between the $p_z$ of C2 and E1 of Co.\\
(ii) Spin-select half-semiconductor: Mn-GY is a spin-select half-semiconductor with 100$\%$ spin polarization at HOMO state. The PDOS of Mn-GY as shown in Fig.\ref{fig7} (c) indicate that it is a semiconductor in both spin channels. In majority spin channel, both the VBM and CBM are derived from the coupling between $p_z$ and $p_y$ states of C2 and E1 states of Mn. In minority spin channel, the VBM mainly derives from the coupling between E2 states of Mn and $p_x$ states of C2, whereas the CBM derives from the coupling between C2 $p_z$ and Mn E1.\\
(iii) Semiconductor-metal transition: The systems realize semiconductor-metal transition, when GDY is adsorbed by Cr, Mn, Fe and Co as shown in Fig.\ref{fig6} (b), (c), (d) and (e), and GY is adsorbed by Cr and Fe, as shown in Fig.\ref{fig7} (b) and (d), respectively. All these systems are spin polarized metals with obvious spin polarization. The spin polarization of Cr-GDY, Mn-GDY, Fe-GDY, Co-GDY, Cr-GY and Fe-GY is 6\%, 18\%, 20\%, 90\%, 53\% and 20\%, respectively. In all these systems, considerable $p_z$ of C1 locate at the Fermi level in both spin channels. Obvious spin polarization of \emph{3d} can be found in these systems. The \emph{$p_z$} of C locate around the Fermi level in the majority spin channel for Cr-GDY, Mn-GDY, Fe-GDY, Co-GDY and Fe-GY, respectively. The \emph{$p_z$} and E1 of Cr locate around the Fermi level in the majority spin channel for Cr-GY. The \emph{$p_z$} of C and  E1 of Mn, Fe, Co locate around the Fermi level in the minority spin channel for Mn-GDY, Co-GDY, Fe-GDY and Fe-GY, respectively, whereas the \emph{$p_z$} of C locate around the Fermi level in the minority spin channel for Cr-GDY and Cr-GY.\\
(iv) Narrow-gap semiconductor: Ni-GDY/GY narrows the band gap in comparison with that of pristine GDY/GY. As shown in Fig.\ref{fig6} (f) and Fig.\ref{fig7} (f), the majority and minority spin electronic states are degenerated for both Ni-GDY and Ni-GY. And both the HOMO and LUMO derive from the $p_z$ states of \emph{sp} hybridized NN carbon atoms and the E1 states of Ni.\\
\section{Conclusion}
\indent In summary, we have performed DFT calculations on the magnetic and electronic properties of single TM atom adsorbed GY/GDY sheet. We find that the adsorption of TM can effectively modulate the magnetic and electronic properties of GY/GDY sheet. V, Cr, Mn, Fe and Co adatoms induce the spin polarization and magnetism for GDY/GY. Most intriguingly, several interesting spintronics properties, such as half-metal and spin-select half-semiconductor, are achieved in the TM-GDY/GY. The modulation of the magnetic and electronic properties of GDY/GY by the adsorption of TM atoms originates from the coupling between TM adatom and GDY/GY and the electron redistribution between intra-atomic orbitals of TM. Our results indicate that TM-GDY/GY systems are potential candidates for spintronics.\\
\begin{acknowledgements}
This work is supported by the Program for New Century Excellent
Talents in University (Grant No. NCET-10-0169), the National Natural
Science Foundation of China (Grant Nos. 10874143 and 10974166), the
Scientific Research Fund of Hunan Provincial Education Department
(Grant No. 10K065), the Hunan Provincial Innovation Foundation for
Postgraduate (Grant No. CX2010B250) and the Specialized Research
Fund for the Doctoral Program of Higher Education (20094301120004).
\end{acknowledgements}


\end{document}